\documentclass[aip,preprint]{revtex4-1}
\usepackage{graphicx}
\usepackage{amssymb,amsmath,amsbsy,amsfonts}
\usepackage{epsfig}
\usepackage[percent]{overpic}
\usepackage{color}

\begin{document}

\title{Heterogeneous dynamics during yielding of glasses: effect of aging}

\author{Gaurav Prakash Shrivastav}
\affiliation{Institut f\"ur Theoretische Physik II, Heinrich-Heine-Universit\"at 
D\"usseldorf, Universit\"atsstr. 1, 40225 D\"usseldorf, Germany}
\author{Pinaki Chaudhuri}
\affiliation{The Institute of Mathematical Sciences, CIT Campus, Taramani, 
Chennai 600 113, India}
\author {J\"urgen Horbach}
\affiliation{Institut f\"ur Theoretische Physik II, Heinrich-Heine-Universit\"at 
D\"usseldorf, Universit\"atsstr. 1, 40225 D\"usseldorf, Germany}

%%%%%%%%%%%%%%%%%%%%%%%%%%%%%%%%%%%%%%%%%%%%%%%%%%%%%%%%%%%%%%%%%%%%%%%%%%%%%%%%%%
%
\begin{abstract}

Molecular dynamics computer simulations of a binary Lennard-Jones
glass under shear are presented. The mechanical response of glassy
states having different thermal histories is investigated by imposing
a wide range of external shear rates, at different temperatures.
The stress-strain relations exhibit an overshoot at a strain of
around 0.1, marking the yielding of the glass sample and the onset
of plastic flow. The amplitude of the overshoot shows a logarithmic
behavior with respect to a dimensionless variable, given by the age
of the sample times the shear rate. Dynamical heterogeneities having
finite lifetimes, in the form of shear bands, are observed as the
glass deforms under shear. By quantifying the spatial fluctuations
of particle mobility, we demonstrate that such shearbanding occurs
only under specific combinations of imposed shear-rate, age of glass
and ambient temperature.

\end{abstract}
%
%%%%%%%%%%%%%%%%%%%%%%%%%%%%%%%%%%%%%%%%%%%%%%%%%%%%%%%%%%%%%%%%%%%%%%%%%%%%%%%%%%
%\newpage

\maketitle

\section{Introduction}
The mechanical properties of amorphous solids have been harnessed
extensively in designing materials which are ubiquitous in our everyday
life. However, a complete microscopic understanding of the mechanisms
leading to the macroscopic response of these materials is still missing.
In order to develop materials with specific functions, it is necessary to
have an improved knowledge of these underlying processes.  This remains
a challenging task.

It is known that the material properties of amorphous solids, such as
colloidal or metallic glasses, depend on their history of production,
e.g.~the cooling rate by which they were quenched from a fluid phase
\cite{glassbook}. This dependence on the history, i.e.~the age of the
amorphous solid, is an important issue in computer simulations of glasses,
especially because the accessible cooling rates in simulations are many
orders of magnitudes larger than those accessible in experiments of real
systems. With respect to the comparison between simulation and experiment,
it is therefore crucial to systematically understand the dependence of
structural and dynamic properties on the age of the glassy solid.

It is thus expected that the response of a glass to an external mechanical
loading is affected by the age of the glass.  If one shears an amorphous
solid under a constant strain rate, it is in general transformed
into a flowing fluid \cite{rodneyrev2011,barratlemaitrerev}. While
at sufficiently high strains, the flowing fluid reaches a steady state
without any memory of the initial unsheared state, the transient response
to the shear is affected by the history of the initial glass state. The
characteristic stress-strain relation of a glassy system, in response
to an externally applied shear rate, exhibits typically a maximum at a
strain of the order of 0.1 \cite{zausch08}.  In numerical simulations
of sheared thermal glasses, the amplitude of this maximum is observed
to depend on the age of the material, typically growing logarithmically
with increasing age \cite{varnik04,robbinsprl05}.

Moreover, the transient response of glasses to an external shear
field is often associated with the occurrence of shear bands,
i.e.~band-like structures with strain or mobility higher than other
regions, observed both in experiments
\cite{schuhrev,mb08,divouxrev15,fs14,vp11,bp10,wilde11,divouxprl10} and
numerical simulations
\cite{vb03,ch13,ir14,gps15,sf06,bailey06,chboc12,ratul-procaccia-12}.  Such
spatially localised structures are seen to emerge after the occurrence
of the stress overshoot, as the stress relaxes to the steady state
value.  Also, the formation of these transient shearbands have been
observed to be influenced by the thermal history of the glassy
state, with states which are obtained by faster cooling being less
susceptible to shearband formation. Further, it has been noted that
such a spatially heterogeneous response is more likely to occur
in the transient regime beyond the stress overshoot, at any given temperature.

The focus of our study are thermal glasses which are often characterized
as simple yield stress fluids, e.g.~colloids, emulsions.  For such
materials, the steady state flow curve (i.e.~stress vs.~imposed shear
rate) is a monotonic function \cite{bp10,nordstrom2010, BBDM15}.. Thus there are no persistent shear-bands,
which would be the case for non-monotonic flow curves \cite{fs14,coussot2010,mb12,ir14}. In the case of
simple yield stress fluids, the transient shearband that emerges are seen
to broaden with time and eventually the entire material is fluidized, with
the timescale of fluidization depending on the imposed shear-rate or stress 
\cite{divouxprl10,divoux12,ch13}. Such
spatio-temporal fluctuations are also visible during steady flow, both
in experiments and simulations \cite{vb03,tsamados10,bp10}. 
However, in this work, our objective is
to characterize the transient spatial heterogeneities, prior to onset of
steady flow. 

The formation of transient shearbands has been addressed within
the scope of various theoretical models. Within the framework of
spatially-resolved fluidity and soft-glass-rheology (SGR) models
\cite{fs14, moorcroft-cates-fielding-11, moorcroft-fielding-13},
the age-dependent spatially heterogeneous
response has been obtained, with the occurrence of the stress-overshoot
under an applied shear-rate being associated with an instability
leading to the formation of these transient heterogeneities. The
model recovers the observation that more pronounced and long-lived
shearbanding occurs for more aged glassy samples. Similarly, Manning et
al.~\cite{manningpre2007,manningpre2009} also observe various transient
heterogeneous states, by analysing a shear-transformation-zone (STZ)
model of glassy materials, which depend upon the initial state of the
system (characterised by an initial effective temperature) and the
imposed shear-rate. Further, they were able to map their results to
those obtained from numerical simulations \cite{shiprl2007}. The same
phenemenology has also been reproduced by other mesoscopic models
\cite{jaglajstat,damienroux2011}.

In this work, we address the question how the combination of ambient
temperature, applied shear-rate and age of the glass affects the
transient response, specifically the observation of spatio-temporal
heterogeneities. This has not been systematically studied in earlier
numerical simulations. Consequently, we also compare our observations
with those from the theoretical models.

To this end, we perform molecular dynamics computer simulations of
the Kob-Andersen binary Lennard-Jones (KABLJ) model \cite{ka94},
a well-studied glass former. Amorphous states are prepared by quenching
a supercooled liquid to different temperatures below the mode coupling
temperature, followed by a relaxation of the sample over a waiting time
$t_{\rm w}$.  Then, the resulting glass samples are sheared with different
shear rates $\dot{\gamma}$.  The onset of plastic flow occurs around
the location of $\sigma^{\rm max}$, i.e.~at a strain $\gamma^{\rm max} =
\dot{\gamma} t^\star \approx 0.1$ (with $t^\star$ the time at which the
maximum is obtained), with the appearance of a peak in the stress-strain
response. We demonstrate that at the different temperatures $T$, the
peak height, $\sigma_{\rm max}$, for all ages and shear-rates, obeys the
functional behavior $C(\dot{\gamma}, T) + A(T) {\rm ln}(\dot{\gamma}t_{\rm
w})$ (with $C$ a function depending on $\dot{\gamma}$ and $T$ and $A$ a
temperature-dependent amplitude). Note that this finding is consistent
with earlier studies \cite{varnik04,robbinsprl05}. Further, as we have
shown recently \cite{gps15}, transient (but long-lived) shear bands
are formed for $\gamma > \gamma^{\rm max}$, provided that shear rate
is sufficiently low.  We quantify the contrast in spatial mobilities
and demonstrate that the extent of spatial heterogeneities is not only
dependent on the age of the glass, but also on the ambient temperature.

The rest of the paper is organized as follows. In Sec.~\ref{sec2} we
describe the KABLJ model and the details of the simulation.  Then, we
present the results for the stress-strain relations and the analysis in
terms of mobility maps in Sec.~\ref{sec3}. Finally, in Sec.~\ref{sec4},
we summarize the results and draw some conclusions.

\section{Model and Methods}
\label{sec2}
We consider a binary mixture of Lennard-Jones (LJ) particles (say A and B)
with 80:20 ratio. This is a well-studied glass former. Particles interact
via LJ potential which is defined as:
\begin{eqnarray}
\label{LJ1}
\textrm{U}^{\textrm{LJ}}_{\alpha\beta}(r) &=& 
\phi_{\alpha\beta}(r)-\phi_{\alpha\beta}(R_{c})-\left(r-R_{c}\right)\left. 
\frac{d\phi_{\alpha\beta}}{dr}\right|_{r=R_{c}},\nonumber\\
\phi_{\alpha\beta}(r) &=& 
4\epsilon_{\alpha\beta}\left[\left(\sigma_{\alpha\beta}/r\right)^{12}-
\left(\sigma_{\alpha\beta}/r\right)^{6}\right]\: r<R_{c},
\end{eqnarray}
where $\alpha, \beta = \textrm{A, B}$. The interaction parameters are
given by $\epsilon_{\textrm{AA}} = 1.0$, $\epsilon_{\textrm{AB}}
= 1.5\epsilon_{\textrm{AA}}$, $\epsilon_{\textrm{BB}} =
0.5\epsilon_{\textrm{AA}}$, $\sigma_{\textrm{AA}} = 1.0$,
$\sigma_{\textrm{AB}} = 0.8\sigma_{\textrm{AA}}$, $\sigma_{\textrm{BB}} =
0.88\sigma_{\textrm{AA}}$, and $R_{c} = 2.5\sigma_{\textrm{AA}}$. Masses
of both type of particles are equal, i.e., $m_{\textrm{A}} =
m_{\textrm{B}} = m$. All quantities are expressed in LJ units in which
the unit of length is $\sigma_{\textrm{AA}}$, energy is expressed
in the units of $\epsilon_{\textrm{AA}}$ and the unit of time is
$\sqrt{{m\sigma_{\textrm{AA}}^{2}}/\epsilon_{\textrm{AA}}}$. More details
about the model and parameters can be found in Ref.~\cite{ka94}.

We perform molecular dynamics (MD) simulation in the $NVT$ ensemble using
the package LAMMPS (``Large-scale Atomic/Molecular Massively Parallel
Simulator'') \cite{plimpton95}.  The simulations are done for the box
geometry having the dimension $20\times20\times80$.  Temperature is kept
constant via a dissipative particle dynamics (DPD) thermostat \cite{sk03}.

Our method for the preparation of the glass samples is as follows: At a
density $\rho=1.2$, we first equilibrate the system at the temperature
$T=0.45$, which is in the super-cooled regime. Then, we quench it to
the target temperatures $T = 0.2, 0.3, 0.4$, below the mode coupling
transition temperature \cite{ka94}.  For exploring the effect of aging on
the mechanical response, we sample glassy states having different ages,
$t_{\rm w} = 10^{2}, 10^{3}, 3\times10^{3}, 10^{4}, 3\times10^{4}$
and $10^{5}$, as the system evolves after the quench to each target temperature.
Using these initial states sampled at different $t_{\rm w}$, we apply shear on
$x$-$z$ plane in the direction of $x$ with different constant strain rates
$\dot{\gamma} = 10^{-2}, 10^{-3}, 3\times10^{-4}, 10^{-4}, 3\times10^{-5}$
and $10^{-5}$. To simulate a bulk glass under shear, we use Lees-Edwards
periodic boundary conditions \cite{le72}.

\section{Results}
\label{sec3}
In order to characterize the mechanical response of the aged amorphous
solids, we measured different structural and dynamical properties, both
at the macroscopic and local scales. We now discuss these observation
in detail.
\subsection{Macroscopic response}
\subsubsection{Stress vs strain}
\begin{figure*}
\includegraphics[width=12cm]{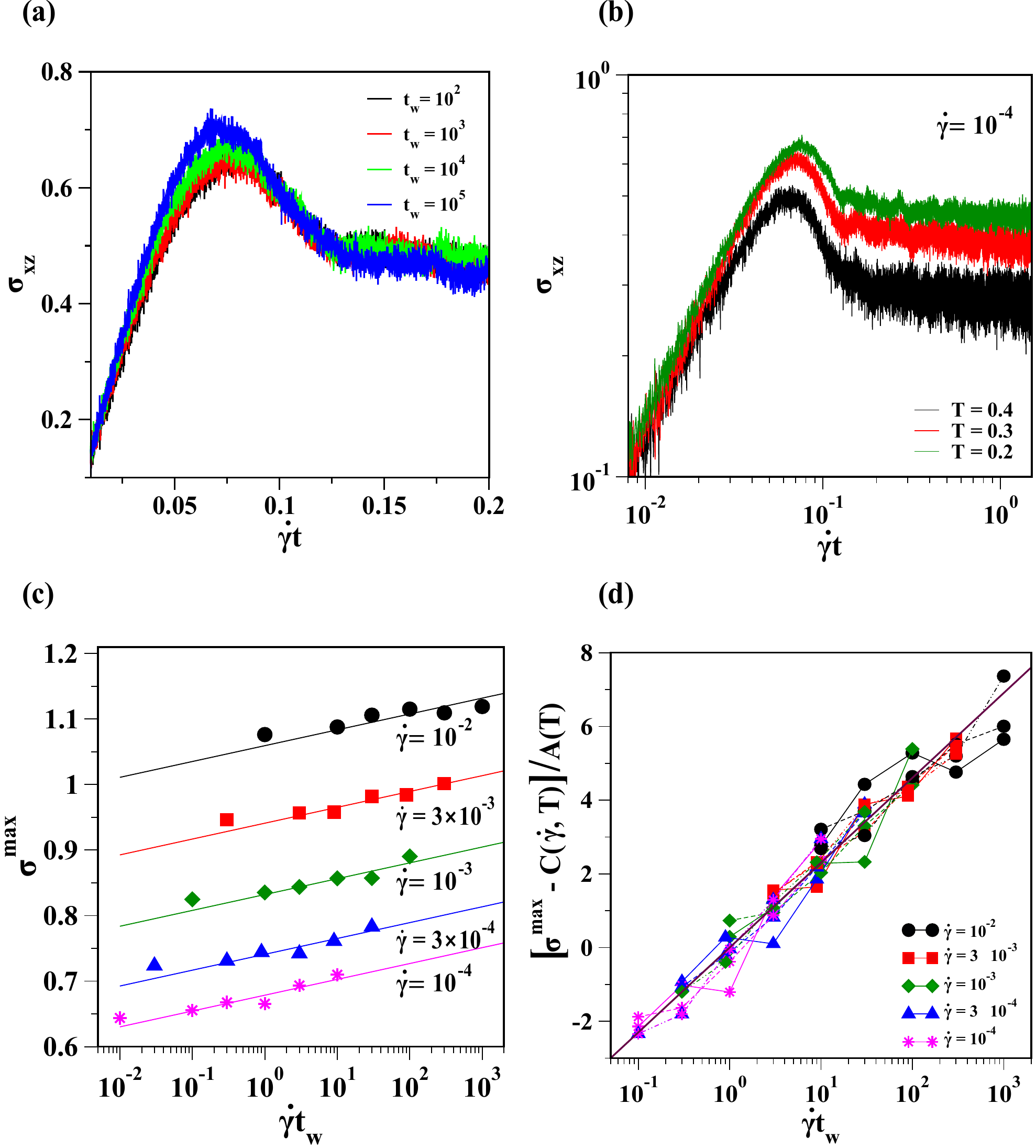}
\caption{\label{fig1} (a) Evolution of macroscopic stress, $\sigma_{\rm
xz}$ with increasing strain, under an imposed $\dot{\gamma}=10^{-4}$ during yielding, for
initial glassy states having different ages $t_{\rm w}$, at $T=0.2$. 
(b) For a fixed shear-rate ($\dot{\gamma}=10^{-4}$) and age ($t_{\rm w}=10^4$),
evolution of $\sigma_{\rm xz}$ with strain at different temperatures,
$T=0.2, 0.3, 0.4$.
(c) The variation of peak height of stress overshoot, $\sigma^{\rm max}$, with age $t_{\rm w}$, for different imposed  $\dot{\gamma}$ at $T=0.2$. Solid lines show the fitting of data with a logarithmic fitting function, $\sigma^{max} = C_{1}(\dot{\gamma}) + A_{1}\ln\left(\dot{\gamma}t\right)$. (d) Scaling of $\sigma^{\rm max}$
with $\ln(\dot{\gamma}t_{\rm w})$ at $T=0.2, 0.3, 0.4$.}
\end{figure*}

When the externally applied shear is imposed on the aging quiescent glass,
the deformation response of the material is characterized by measuring
the stress, $\sigma_{\rm xz}$, generated in the system with increasing
strain ($\dot{\gamma}t$). In Fig.~\ref{fig1}(a), we show how the stress
evolves when we impose $\dot{\gamma}=10^{-4}$ on initial states sampled
from different ages, at $T=0.2$ . We observe the characteristic stress
overshoot \cite{varnik04}, with the peak height ($\sigma^{\rm max}$)
increasing for larger $t_{\rm w}$. Across temperatures, within the glassy
regime, this characteristic response does not change. However, as
expected the material becomes less rigid with increasing temperature.
For example,  for a fixed age, with increasing temperature, the steady
state value of $\sigma_{\rm xz}$ decreases, but still one observes the
stress-overshoot, albeit with a lesser value of $\sigma^{\rm max}$;
see Fig.~\ref{fig1}(b) for the corresponding data at $T=0.4,0.3,0.2$
for an imposed $\dot{\gamma}=10^{-4}$.

In Fig.~\ref{fig1}(c), we illustrate the variation of $\sigma^{\rm max}$
with age, for a wide range of $\dot{\gamma}$ at $T=0.2$. Thereafter,
we demonstrate that at each temperature, the data for $\sigma^{\rm max}$
can be collapsed onto a master curve using the relation
\begin{equation}
\sigma^{\rm max} = C(\dot{\gamma}, T) 
+ A(T) {\rm ln}(\dot{\gamma}t_{\rm w}),
\end{equation}
with $A(T)$ a temperature-dependent amplitude that is independent of
$\dot{\gamma}$ and $t_{\rm w}$ and $C(\dot{\gamma},T)$ a function that
solely depends on $\dot{\gamma}$ and $T$.

Such a logarithmic relationship is motivated by Ree-Eyring's viscosity
theory, modified appropriately to take into account the role of the
sample's age, $t_{\rm w}$, as noted by Varnik et al.~\cite{varnik04}. A similar
dependence was also proposed by Rottler et al.~\cite{robbinsprl05}
for the same binary LJ mixture subjected to uniaxial strain at a
different state-point. Thus our scaling results are consistent with
earlier observations. We note that such a logarithmic dependence
is not observed in experiments involving other yield stress fluids like
colloidal pastes \cite{derec2003} or carbopol \cite{divouxsoftmatter2011}
where a power-law increase is observed, which is also captured in the
numerical simulations of model gels \cite{parksoft13,whittle97}. Thus,
it seems that the response of dense thermal glasses are different from
the low density materials with more complex structures.

Having observed the scale of the stress overshoot for various ages,
imposed shear-rates and ambient temperatures, we will later explore
whether such observations necessarily lead to the occurrence of transient
shearbands.
\subsubsection{Potential Energy}
\begin{figure*}
\includegraphics[width=16cm]{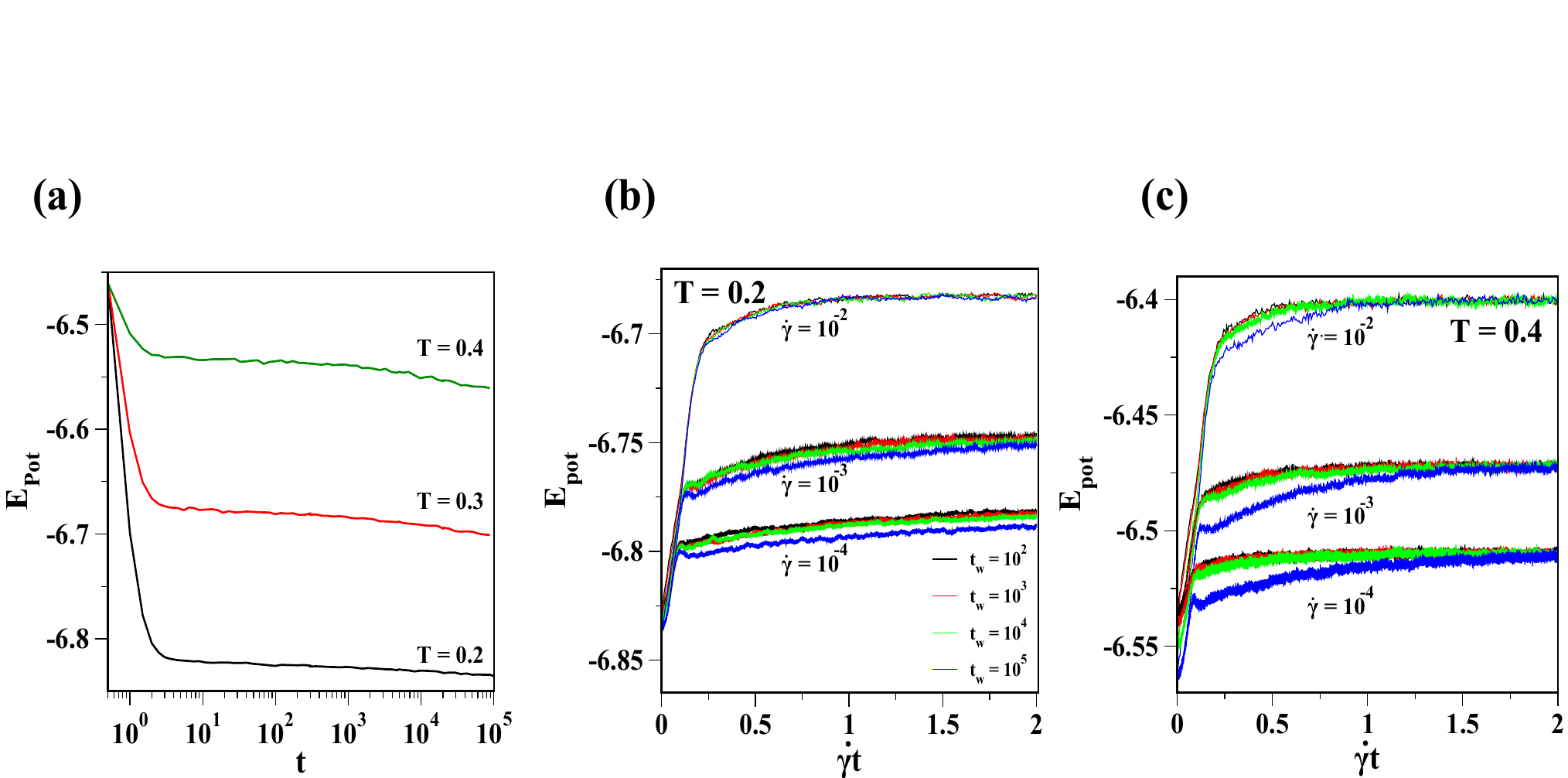}
\caption{\label{fig2} (a) Evolution of potential energy ($E_{\rm Pot}$) of the
quiescent glass during aging for three different temperatures, showing typical logarithmic decay. (b) For $\dot{\gamma}=10^{-2}, 10^{-3}, 10^{-4}$, increase of potential energy with strain ($\dot{\gamma}t$) for different $t_{\rm w}$ of the quiescent glass at $T=0.2$. (c) Similar plot as in (b) {for T=0.4}}
\end{figure*}

To have a measure of the structural changes in the system, both during
the aging process and the onset of flow, we monitor the potential energy
($E_{\rm {Pot}}$) of the system.  In Fig.~\ref{fig2}(a), we show how the potential
energy decreases when the system is quenched from the supercooled state
at $T=0.45$ to the lower temperatures $T=0.2, 0.3, 0.4$. After the
fast decrease during the initial aging regime, the potential energy
decreases slowly, logarithmically depending on the age $t_{\rm w}$ of the
system, as is typically observed during aging.

When the external shear is imposed on the system, the potential energy
increases, with the steady state value depending upon the magnitude
of the imposed shear-rate; see Fig.~\ref{fig2}(b),(c) for $T=0.2,
0.4$, respectively. Further, we check whether the potential energy of
the system measured during the transient regime, prior to the onset of
steady flow, is dependent on the initial age of the quiescent glass. As
can be seen in Fig.~\ref{fig2}(b), (c), for both the temperatures, the
dependence on $t_{\rm w}$ becomes more visible with decreasing shear-rate. This
thus indicates that the system evolves through different intermediate
structures, dependent on the initial state having different $t_{\rm w}$,
before steady states structures start getting explored. We now explore
how these age-dependent transient structures are dynamically different. We
also note that for the applied shear-rates, the approach to steady state
seems faster for $T=0.4$ than compared to $T=0.2$. While in the former
case, at long strain, the steady state potential energy, under applied
shear, matches at large strains, this is not the case for $T=0.2$. This
combination of temperature, shear-rate and age of the sample will be
further explored in later sections.
\subsection{Single particle dynamics}
\subsubsection{Average MSD}
\begin{figure*}
\includegraphics[width=14cm]{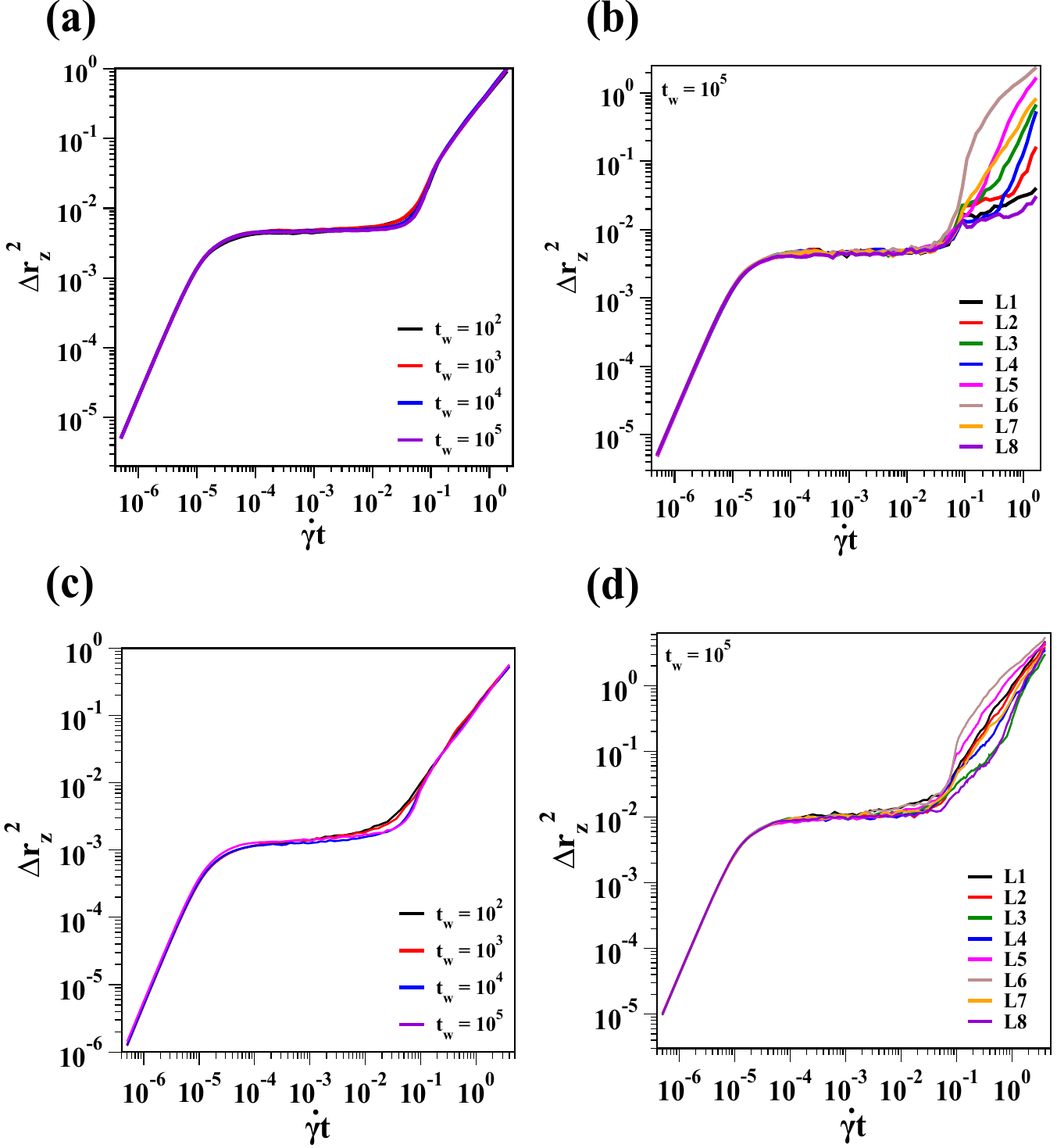}
\caption{\label{fig3} (Top) $T=0.2$: (a) Variation in  $z$-component
single particle MSD, $\Delta r_{z}^{2}$, with age for a fixed
imposed $\dot{\gamma}=10^{-4}$.  (b) Spatial variation in MSD for
$\dot{\gamma}=10^{-4}$ and $t_{\rm w}=10^5$.  (Bottom) $T=0.4$ (c) Variation
in $\Delta r_{z}^{2}$, with $t_{\rm w}$ for imposed $\dot{\gamma}=10^{-4}$. (d)
Spatial variation in MSD for $\dot{\gamma}=10^{-4}$ and $t_{\rm w}=10^5$.}
\end{figure*}

In order to characterise the microscopic dynamical response of the glass,
when yielding under shear, we monitor the single-particle dynamics.
This is quantified by measuring the non-affine mean-squared
displacement (MSD), $\Delta r_{z}^{2}$, in the direction transverse to
the applied shear; the data is shown for $\dot{\gamma}=10^{-4}$, at
$T=0.2$ (Fig.~\ref{fig3}(a)) and $T=0.4$ (in Fig.~\ref{fig3}(c)). At
both temperatures, during early times, particles exhibit ballistic
motion, which is then followed by a caging regime.  Subsequently, the
onset of flow is marked by a super-linear regime, and eventually diffusive
dynamics is observed. The onset of the super-linear or ``super-diffusive''
regime occurs around the stress overshoot in the stress-strain curve,
when the stress, building up in the system, is released via the breaking of
local cages leading to the subsequent diffusive motion of the particles
\cite{zausch08, koumakisprl12}. We check how the initial age of the glassy state
influences the measured $\Delta r_{z}^{2}$; the corresponding data is
plotted in Fig.~\ref{fig3}(a), (c) for the two temperatures.  At both
temperatures, both in the initial response and in the steady state,
there is no visible difference in the measured MSD, for the samples
having different ages. Only during the transient regime, where the
onset of super-diffusive dynamics occurs, we observe variation with
age of the quenched glass, with the effect more visible for $T=0.4$;
see Fig.~\ref{fig3}(c).  We also note that the onset timescale depends
upon age, shifting to longer timescales with increasing $t_{\rm w}$, as is
clearly visible.
\subsubsection{Spatially resolved MSD}
To further probe the spatial features of the local dynamics during
yielding, we divide the simulation box into eight layers of thickness $10\sigma_{\rm AA}$ along the
$z$-direction. For each of these layers, we compute the averaged MSD
for the particles populating it at $t=0$, i.e.~before shear is imposed.
In Fig.~\ref{fig3}(b), (d), we show the corresponding data, at $T=0.2,
0.4$, respectively, for an imposed $\dot{\gamma}=10^{-4}$ and age
of $t_{\rm w}=10^5$. We see that there is variation in the dynamics across
the different layers, the variation being greater at $T=0.2$ than at
$T=0.4$. While for $T=0.4$, eventually, the long time dynamics seems to
converge to the diffusive limit, within the timescale of observation,
that is not the case for $T=0.2$. In the latter case, while in one case,
we see the onset of diffusive dynamics at long times, in some of the
other slices, the dynamics continues to be sub-diffusive. To summarise,
this demonstrates that, during yielding of the glass, at any temperature,
dynamics is heterogeneous. However, the extent of heterogeneity and
persistence increases with decreasing temperature.  We will quantify
that in more detail in the subsequent sections.
\subsubsection{MSD maps}
\begin{figure*}
\includegraphics[scale=0.7, clip=true]{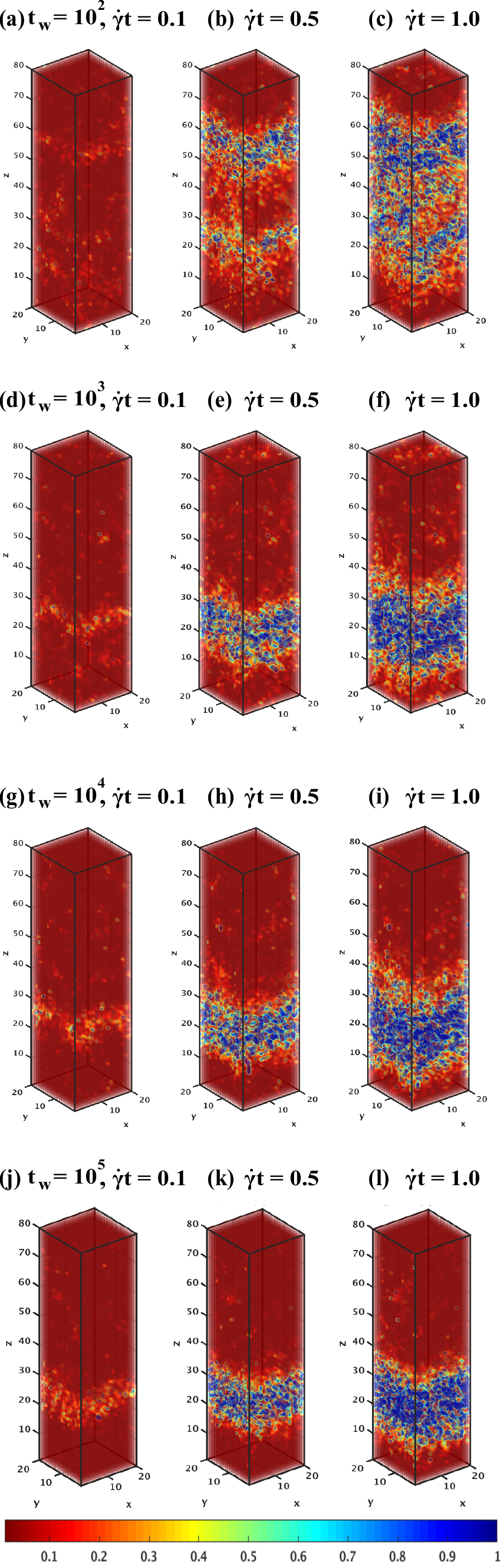}
\caption{\label{fig4} At $T=0.2$, for imposed shear 
rate of $\dot{\gamma}=10^{-4}$,
MSD maps at different strains, during the yielding of the glass of
different ages $t_w=10^{2} ({\rm top}),t_w=10^{3} ({\rm second}),
t_{\rm w}=10^{4} ({\rm third}), t_{\rm w}=10^{5} ({\rm bottom})$.}
\end{figure*}

\begin{figure*}
\includegraphics[scale=0.7, clip=true]{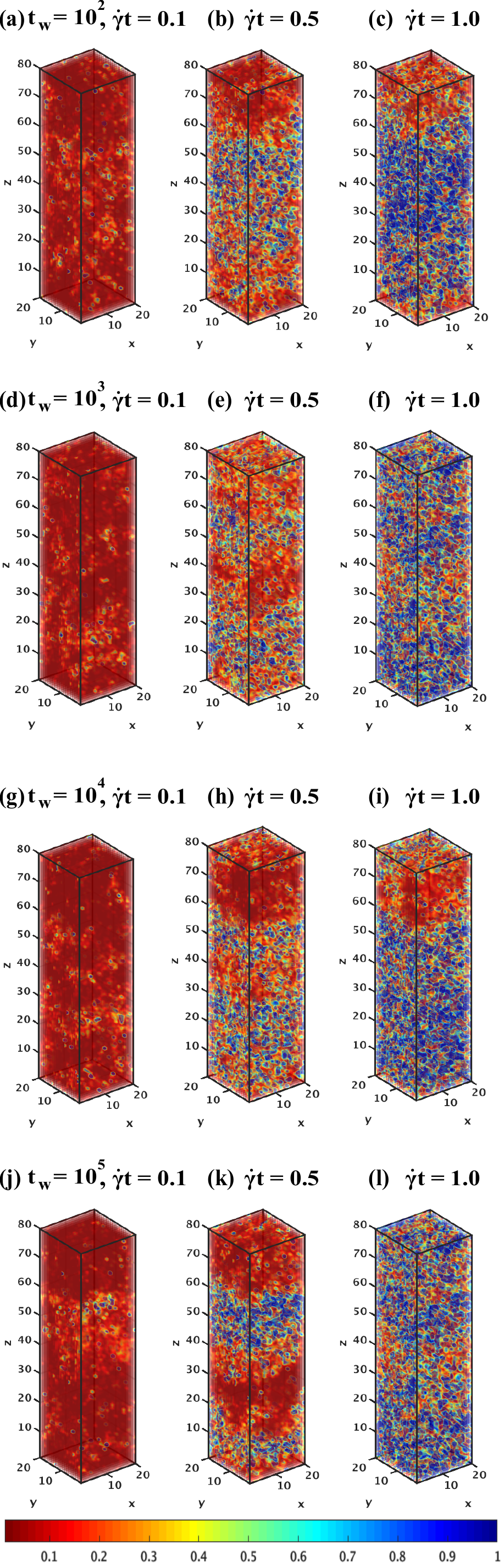}
\caption{\label{fig4b} At $T=0.4$, 
for imposed shear rate of $\dot{\gamma}=10^{-4}$,
MSD maps at different strains, during the yielding of the glass of
different ages $t_{\rm w}=10^{2} ({\rm top}),t_{\rm w}=10^{3} ({\rm second}),
t_{rm w}=10^{4} ({\rm third}), t_{\rm w}=10^{5} ({\rm bottom})$.}
\end{figure*}

To visualise the dynamical heterogeneities, we construct three-dimensional
maps of local MSD \cite{ch13}. These maps are constructed in the
following manner.  We divide the simulation box into small cubic
sub-boxes having linear size of $\sigma_{\textrm{AA}}$.  At any time
$t$, we calculate the average MSD of  the particles located in each
sub-box at $t=0$ (unsheared glassy state), which provides a cumulative
picture of how yielding proceeds locally starting from the quiescent
glass. As discussed earlier, we measure the $z$-component of MSD of each
particle. The evolution of the three-dimensional maps with increasing
strain, is shown in Fig.~\ref{fig4} and Fig.~\ref{fig4b}, for different
ages of the glassy state at $T=0.2$ and $0.4$, respectively, corresponding
to an imposed shear rate of $\dot{\gamma}=10^{-4}$.

First, we focus on the situation at the lower temperature ($T=0.2$);
see Fig.~\ref{fig4}. At early strain ($\dot{\gamma}=0.1$), we observe
localised ``hot spots'' of large mobilities \cite{clement12,sentja15}.
At a strain of $0.5$, more such local regions of faster dynamics have
emerged. In the case of well aged initial state (e.g.~$t_{\rm w}=10^5$), we see
that the hot spots are organised in the form of a shear band. On the other
hand, for a younger initial state, i.e.~$t_{\rm w}=10^2$, these hot spots are
much more dispersed. At a later strain ($\dot{\gamma}=1$), this difference
is further amplified.  For the younger sample, large regions of the system
have been fluidized, whereas for the older system, enhanced mobility
is still localised in the form of a shear band. Thus, the age of the
initial glassy state, does influence the spatio-temporal organisation of
regions of increased mobility.  This is consistent with earlier findings
in numerical simulations using different model systems \cite{sf06}
or predictions from theoretical models \cite{moorcroft-cates-fielding-11}.

We now contrast this with the situation at the higher temperature
($T=0.4$); see Fig.~\ref{fig4b}. At early strain ($\dot{\gamma}=0.1$),
we observe the hot spots, for all $t_{\rm w}$; however, there are more in number
as compared to the case of $T=0.2$, for any $t_{\rm w}$. With increased strain
$\dot{\gamma}=0.5$, we see that the hot spots have proliferated and cover
almost the entire domain, including for the most aged sample. Thus,
access to increased thermal fluctuations at higher temperatures lead
to the tendency for a more homogenised and faster fluidisation of the
glassy state under applied shear.

We will now carry out a more quantitative characterisation of this
spatio-temporal response of the glassy states, sampled from different
temperatures, and aged over different timescales.
\subsubsection{Spatial profile of local mobility and fluctuations}
So far, by spatially resolving the single particle MSDs, we demonstrated
the existence of heterogeneous dynamics during yielding of the glass
and how that depends on the temperature of the glassy state.

\begin{figure*}
\includegraphics[scale=0.7,clip=true]{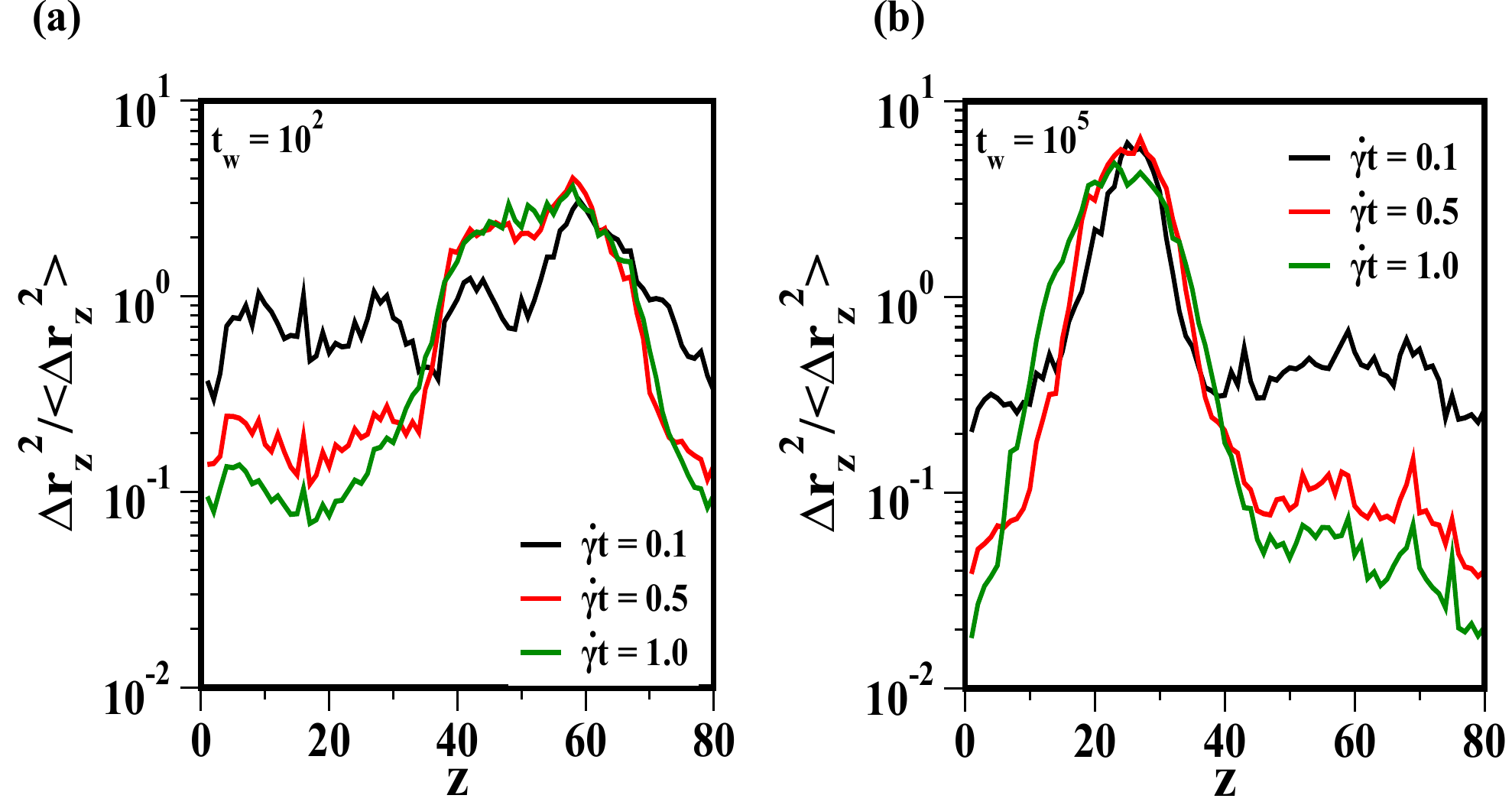}
\caption{\label{fig6} Evolution of spatial profiles of local MSD,
$\Delta r^{2}_{z}(z)/\langle{\Delta r^{2}_{z}}\rangle$, with strain for
(a) $t_w=10^2$, (b) $t_w=10^5$, under an imposed shear of
$\dot{\gamma}=10^{-4}$.}
\end{figure*}

Using single particle MSDs, we construct the instantaneous spatial
profile $\Delta r^{2}_{z}\left(z\right)/\langle{\Delta r^{2}_{z}}\rangle$, which gives us
a measure of fluctuations in local dynamics. The evolution of these
profiles, with increasing strain, is shown in Fig.~\ref{fig6}, for two
different ages of the glass, viz.~$t_{\rm w}=10^2, 10^5$, under an imposed
shear-rate of $\dot{\gamma}=10^{-4}$, at $T=0.2$.  We observe that the
spatial fluctuations are much larger and more persistent in the older
sample, consistent with the qualitative discussion, above, involving
MSD maps.

Using these spatial profiles, we can now quantify the
contrast in local mobilities in the deforming glass, under
applied shear, and compare the response across temperatures.
In order to do that, we compute the fluctuations in local MSD,
$\chi=(\Delta r^{2}_{z}\left(z\right)-\langle{\Delta r^{2}_{z}}\rangle)^2/\langle{\Delta r^{2}_{z}}\rangle^2$
with increasing strain $\dot{\gamma}t$.  Such a quantity measures the
extent of fluctuations in average mobilities, as characterised by the
local MSD, across the regions parallel to the direction of flow. Defined
in this manner, $\chi$ would therefore easily capture the contrast in the
dynamics during the formation of shear-bands. To note, such a quantity
is different from the dynamical susceptibility, $\chi_4$, which is a
measure of fluctuations in single particle dynamics, measured within
the steady-state ensemble. Here, we are concerned with the fluctuations
across regions during the onset of flow from a quiescent state.

\begin{figure*}[h]
\includegraphics[scale=0.8, clip=true]{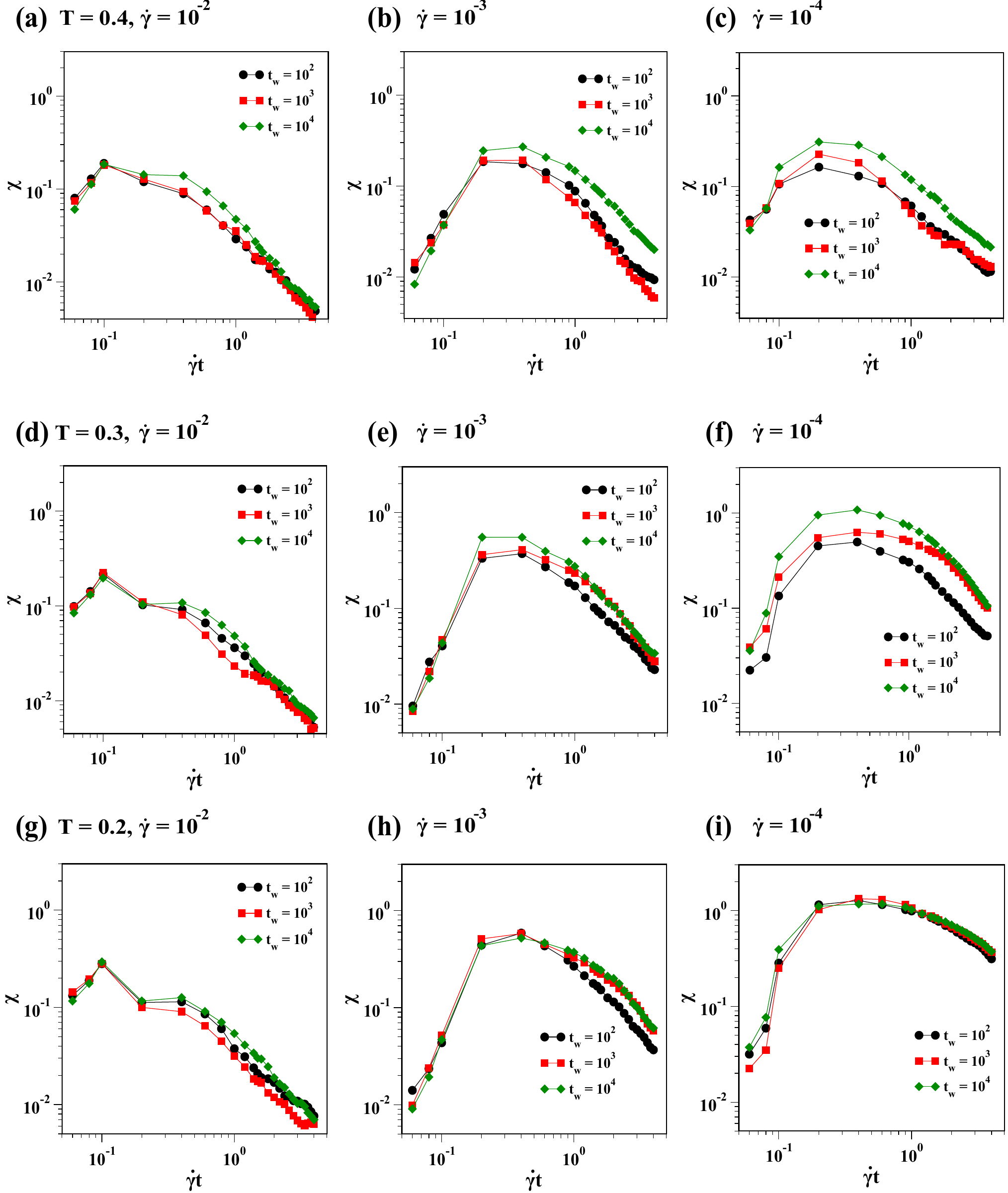}
\caption{\label{fig7}Variation of the fluctuations in local MSD, $\chi$,
with strain for different $t_{\rm w}$, at $\dot{\gamma}=10^{-2}$, $10^{-3}$,
${10^{-4}}$ (left to right), for glassy states sampled at temperatures
$T=0.4$ (top row), 0.3 (middle row), 0.2 (bottom row).}
\end{figure*}

The corresponding data, for temperatures $T=0.2, 0.3, 0.4$, across ages
$t_{\rm w}=10^2, 10^3, 10^4, 10^5$, for a range of imposed shear-rates,
viz.~$\dot{\gamma}=10^{-2}, 10^{-3}, {10^{-4}}$, are shown in
Fig.~\ref{fig7}. Typically, $\chi$ behaves non-monotonically with
increasing strain, with the maximal spatial fluctuations ($\chi^{\rm
max}$) occuring at a strain which corresponds to the super-linear regime
in the MSD (see Fig.~\ref{fig3}); i.e., once yielding of the glass
has occurred.  Further, for a fixed $\dot{\gamma}$, the magnitude of
$\chi^{\rm max}$ increases with age, i.e.~the contrast in mobilities
across a sample becomes largest for a more aged sample. Importantly,
we also note that $\chi^{\rm max}$ also starts increasing when we
shear the glasses with smaller and smaller shear-rates; e.g.~see
Fig.~\ref{fig7}(a)-(c), for the variation of $\chi^{\rm max}$ with
age and imposed shear-rate. With decreasing temperature, this spatial
fluctuation increases even further.

\begin{figure*}
\includegraphics[scale=0.8, clip=true]{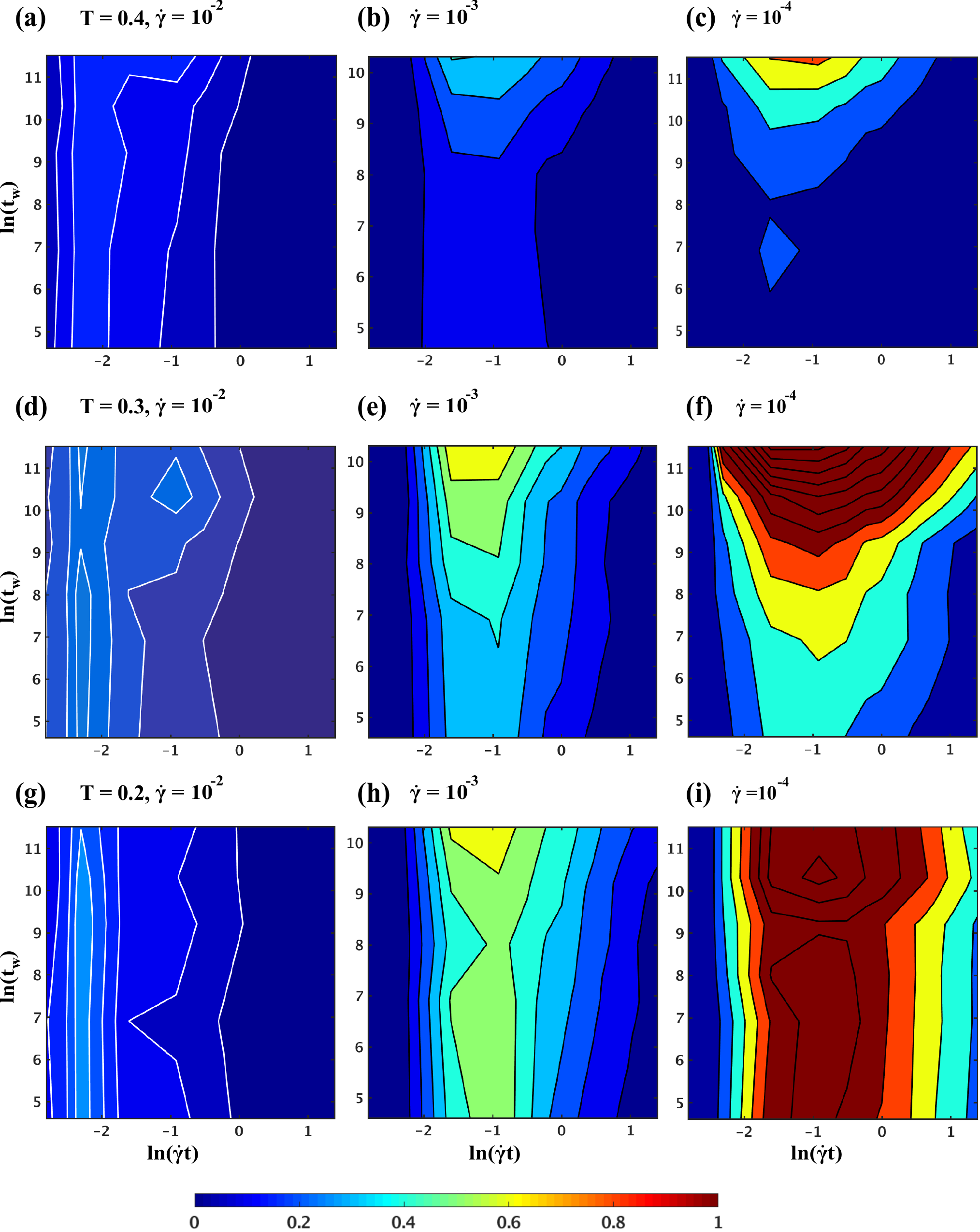}
\caption{\label{fig8}Contour maps of $\chi(t_{\rm w}, \dot{\gamma}t$), for
$T=0.4$ (top panel), $T=0.3$ (middle panel), $T=0.2$ (bottom panel),
under imposed shear-rates of $\dot{\gamma}=10^{-2}, 10^{-3}, 10^{-4}$
(from left to right in each panel).}
\end{figure*}

To get a more complete picture of the variation of these spatial
fluctuations, across age and temperature, we plot the corresponding
contour maps for $\chi(t_{\rm w},\dot{\gamma})$; these are shown in
Fig.~\ref{fig8}. From the contour maps, one can infer how strong and
persistent the spatial fluctuations are with varying temperature, age
and imposed shear-rate.  It is evident that for large shear-rates
($\dot{\gamma}=10^{-2}$), these fluctuations are negligible at
any age of the glassy state sampled at any of the temperatures; see
Fig.~\ref{fig8}(a),(d),(h). On the other hand, at smaller shear-rates
($\dot{\gamma}=10^{-4}$), large fluctuations are visible at all
temperatures. However, the age of the sample then becomes a factor in
determining the degree of fluctuations and persistence with increasing
strain.  At a lower temperature, $T=0.2$, these are prominent over the
range of ages we investigated and also persistent over large strains;
see Fig.~\ref{fig8}(i). On the other hand, at higher temperatures, $T=0.4$
(Fig.~\ref{fig8}(c)), that scale of fluctuations is only slightly visible
at very large ages over a small strain window.  The range of ages and
strain-interval over which similar fluctuations are seen broadens out at
an intermediate temperature of $T=0.3$ (Fig.~\ref{fig8}(f)) and, even more
extensively, at $T=0.2$. This trend is also evident at an intermediate
shear-rate of $\dot{\gamma}=10^{-3}$, with the scale of fluctuations
being less but visible (Fig.~\ref{fig8}(h)); howevever, it is important
to note that the fluctuations are nearly negligible at the higher
temperature $T=0.4$ at this shear-rate (Fig.~\ref{fig8}(b)).  Thus, we
demonstrate, that the scale of transient dynamical heterogeneities depend
on a combination of temperature of the glass, its age and the imposed
shear-rate. Since the quantity we compute, $\chi(t_{\rm w},\dot{\gamma})$,
measures the contrast in mobilities in layers parallel to the flow
direction, the conclusions drawn from the contour maps allow us to infer
that the propensity to shear-band during yielding is determined on how
rigid the glass is and how small/large the applied shear-rate is.
\subsubsection{Time evolution of shear-bands}
Finally, we study how a transient shear-band, once formed, spatially
evolves with time. Such analysis can only be done for the case where a
well-defined shear-band can be identified. Thus, we focus at the case of
$T=0.2$ and small shear-rate ($\dot{\gamma}=10^{-4}$), where we observe
transient shear-bands for a range of $t_{\rm w}$.

\begin{figure*}[htb]
\includegraphics[scale=0.75,clip=true]{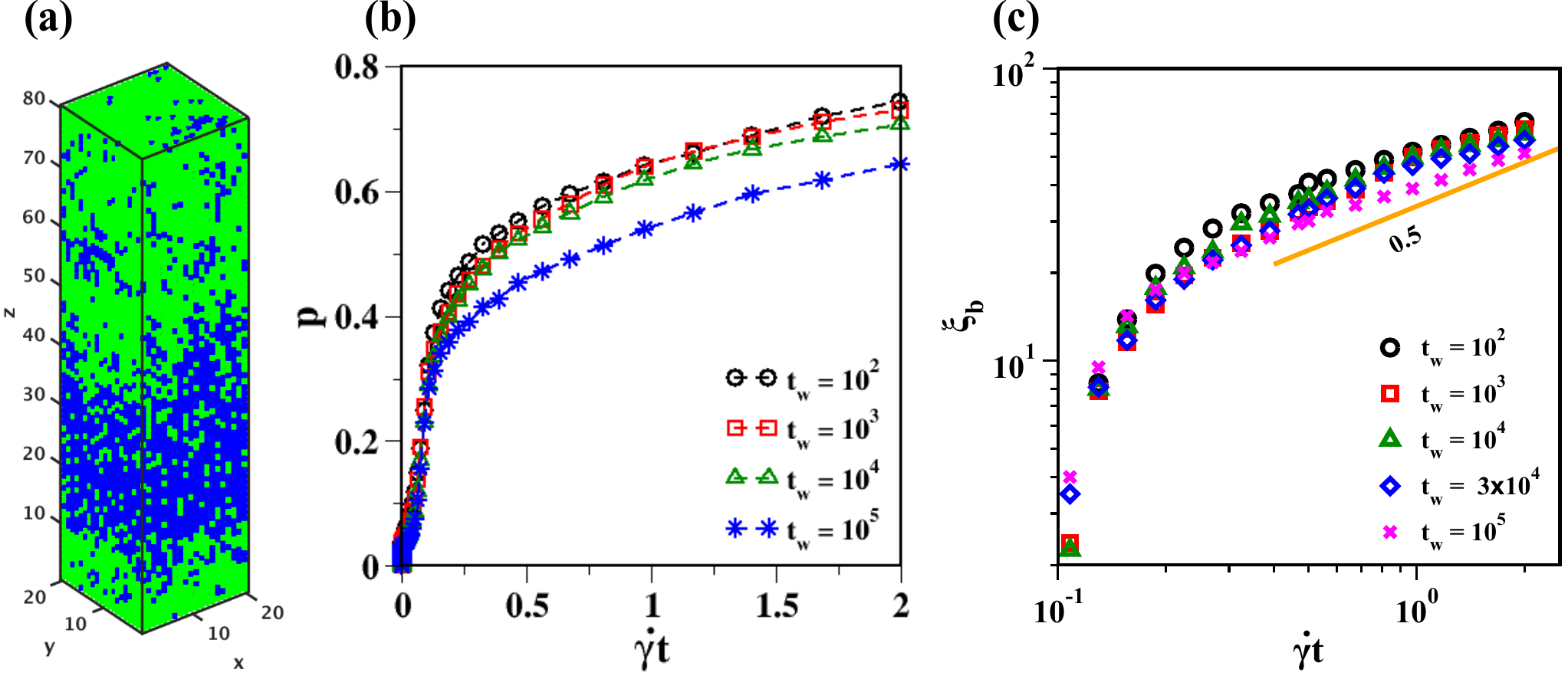}
\caption{\label{fig9} (a) Map of local mobility $\psi$ of a sample
of an age $t_{\rm w} = 10^4$ for $T=0.2$, $\dot{\gamma}t=0.5$.
Mobile regions are marked in blue while immobile regions are marked
in green.  (b) Growth of fraction of mobile regions with increasing
strain, for imposed  $\dot{\gamma}=10^{-4}$ and different ages.
(c) Increasing width of the shearband, $\xi_b$, with strain for
different $t_{\rm w}$, at imposed $\dot{\gamma}=10^{-4}$.}
\end{figure*}

In order to further quantify and characterise the spatio-temporal
evolution of the shear-band, we define a region to be mobile or not,
by setting a threshold $\mu_{th}=0.02$ on the local $\Delta r^{2}_{z}$. As
can be seen in Fig.~\ref{fig3}(a), such a choice of $\mu_{th}$ is larger
than the plateau value in the MSD and thus corresponds to motions beyond
cage-breaking. We then define the local mobility $\psi$ as
\begin{eqnarray}
\label{orderparam}
\psi = \begin{cases} 1 &\mbox{if } \mu \ge \mu_{th}, \\
                     0 & \mbox{otherwise},\end{cases}
\end{eqnarray}
where $\mu$ is the average MSD of particles in a sub-box. Following
this convention, we digitize the whole system into mobile and immobile
regions. An example of the mobility map, thus constructed, is shown
in Fig.~\ref{fig9}(a).

We quantify the extent of fluidisation in the system by measuring the
fraction of mobile cells, $p$, at any given instant. The evolution of $p$
as a function of strain, for an imposed shear of $\dot{\gamma}=10^{-4}$,
is shown in Fig.~\ref{fig9}(b). We observe that independent of age
($t_{\rm w}$), the sudden initial increase of $p$ occurs nearly at the same
strain. However, the subsequent increase in $p$ does depend upon the age
and thus, the fraction of system which is mobilised at $\dot{\gamma}t=0.2$
is substantially larger for $t_{\rm w}=10^2$ as compared to $t_{\rm w}=10^5$.

Further, we try to identify the spatial organisation of these mobile
cells. By locating contiguous layers of mobile cells, we identify the
formation of the shear band and, thereafter, by marking the interfaces
of this band, we measure how the band-width, $\xi_{\textrm{b}}$, evolves
with time. For different ages of the initial glassy state, this time
evolution is shown in Fig.~\ref{fig9}(c), for a fixed
$\dot{\gamma}=10^{-4}$. We see that $\xi_{\textrm{b}}$ initially grows
quickly and then eventually it reaches a regime where the data can be
fitted with $\xi_{\textrm{b}} \sim t^{1/2}$, implying that the propagating
interface of the shear band has a diffusive motion. This diffusive regime
is not extensively dependent on $t_{\rm w}$, although samples having largest
$t_{\rm w}$ do seem to display a slower motion of the spreading interface. 

\section{Summary and conclusions}
\label{sec4}

To summarize, we have used numerical simulations of a model glass
former, to probe the mechanical response of amorphous solids, having
different aging histories, by imposing a wide range of external
shear rates. The response is studied at different temperatures below
the mode coupling temperature, in order to ascertain how thermal
fluctuations affect the spatio-temporal response. We underline,
here, that the numerical simulations are done by integrating the
Newton's equations of motions using Lees-Edwards periodic boundary
conditions and the local DPD thermostat. This, thus, allows for
unhindered spatio-temporal fluctuations as the glassy state responds
to the applied shear, without introducing any biases or suppressing
any fluctuations, which would be the case if one were using walls
\cite{vb03} or integrating the SLLOD equations along with some
global thermostat (like Nose-Hoover) \cite{shiprl2007}. Therefore,
qualitatively, the nature of dynamical heterogeneities observed
could be different from the earlier works.

The macroscopic response of the glass is characterized by the
occurrence of a overshoot in the stress-strain curves, observed for
the range of shear-rates and ages that we have studied, at the
different temperatures. Consistent with earlier works, we find that,
at each temperature, the peak height can be scaled on to a master-curve
as a function of $\dot{\gamma}t_{\rm w}$. The corresponding potential
energies of the system, measured during the deformation process,
show a dependence on age with decreasing shear rate. This implies
that the structures visited during the transient regime, prior to
the onset of steady flow, start becoming different, depending upon
the initial history. We also observe that for lower temperatures,
it takes longer for potential energies to reach the steady state
value at smaller shear-rates, reflecting the role of thermal
fluctuations in affecting the duration of the transient regime for
these parameters.

This is further revealed, by measuring dynamical quantities at the
microscopic scale. Our probe of choice is the time evolution of non-affine
mean squared displacement (MSD) of the particles relative to the initial
quiescent state.  Spatial profiles of the MSD show that for a fixed
imposed shear and age, the transient dynamics is more heterogeneous
and more long-lived at lower temperatures. This is illustrated by
constucting sptatially resolved three-dimensional maps which provide
a handle to visualize the extent of dynamical heterogeneity as the
glassy states respond to the imposed shear. Such maps demonstrate that
the heterogeneities are spatially localised in the form of shearbands,
for old enough systems at low temperatures.

We quantify the extent of dynamical heterogeneity organised in the
form of shear-bands, by measuring the degree of fluctuations, $\chi$,
in local mobilities across regions oriented parallel to the direction
of imposed flow. Thus, large values in $\chi$ reflect more propensity
of the system to exhibit shear-banding. We do such measurements by
scanning across entire range of shear-rates and thermal histories
explored by us. We observe that the fluctuations are maximum around
the yielding strain and the peak in fluctuations increase with
decreasing shear-rate as well as larger ages. Further by constructing
contour maps of $\chi (t_{\rm w}, \dot{\gamma}t)$ for varying
shear-rates at different temperatures, we demonstrate that the
emergence of significant shear-banding depends upon the combination
of temperature, age and imposed shear-rate. The contrast in local
mobilites is most prominent at low shear-rates and low temperatures,
where the age of the system does not seem to matter, over the range
explored by us. With increasing temperature, the age of the sample
starts to determine whether transient shear-banding will be visible
or not. For even higher temperatures, one has to go to lower
shear-rates and even older systems in order for such largely localized
heteroegeneities to be visible. Also, in the cases where shear-bands
can be identified, we observe that the spreading of the shear-band
across the system seems to depend on age, albeit weakly.

Thus, our study shows that the transient shearbanding, with sharply
contrasting mobilities across regions, only occur for specific
combinations of temperature, shear-rate and age, as discussed above.
Note that, for most of the glassy samples that we study, the
macroscopic response under an imposed strain-rate is characterized
by the stress overshoot. Yet, the transient shear-bands are only
visible in only a subset of the cases we study.  Thus, the occurrence
of a stress overshoot does not necessarily lead to the transient
shearbands, quantified via the mobilities relative to the initial
quiescent state, as predicted by the spatially resolved fluidity
and SGR models \cite{fs14, moorcroft-cates-fielding-11, moorcroft-fielding-13}. 
On the other hand, as predicted by these and other
models \cite{manningpre2007,manningpre2009, jaglajstat,damienroux2011}, 
we do observe that the propensity to shearband is more in
the case of increasingly aged samples.  However, the choice of
disorder distributions in these models (e.g. distribution of yielding
thresholds), with changing age, remains arbitrary, and it would be 
useful to obtain these as inputs from atomistic simulations.

The transient shear-banding observed in our numerical simulations
as well as other, bears resemblance to dynamical heterogeneities
observed in thermal glasses \cite{ludormp11}, except that in the
case of a sheared glass, strong anisotropies come into play leading
to the spatial localisation. In the case of supercooled liquids,
the scale of dynamical heterogeneities increases as the temperature
is decreased towards the putative glass transition temperature.
Similarly, in the case of applied shear, the heterogeneous dynamics
is enhanced as one approaches vanishingly small shear-rates, with
the yield stress being identified as a critical threshold
\cite{bocquet-prl-2009, divoux12, chenprl2016}. Also, it is possible
that the regime of imposed shear-rates over which the critical
threshold influences flow \cite{ludo-jpcm-03, chboc12} could change
with temperature, with decreasing thermal fluctuations extending
this regime over longer scales. This would rationalise why with
increasing temperatures, the contrast in heterogeneities progressively
decreases, for a fixed shear-rate. Over and above, the aging effects
are more prominent at higher temperatures, with the glass remembering
the history from where it was quenched over longer aging timescales
\cite{ludo-jpcm-03, ludormp11}.  At low temperature, thus, structural
arrest and therefore rigidity emerges at shorter timescale.  Hence,
the mechanical response of the glass, having the same age and same
imposed shear-rate, would be different across temperatures.  This
interplay of different timescales therefore brings about the complex
response as a function of all these control parameters.  Eventually,
it would be useful to quantify the exact ranges in these control
parameters, where we expect transient but prominent shearbanding
to be visible.

The transient dynamical heterogeneities that we track and analyse
are similar to measurements done via confocal microscopy in colloidal
glasses \cite{vp11} or scattering measurements in granular systems
\cite{clement12}.  This is different from using velocity profiles
as a diagnostic tool, which capture plasticity over short timescales.
Further, in our case, because of periodic boundary conditions, the
shear bands can emerge anywhere in the system, unlike many cases
where initial mobile regions happen to occur near confining surfaces
\cite{divouxsoftmatter2011, vp11}.  However, similar to experiments,
we do observe the propagation of the mobile front to eventually
fluidize the system.
{\bf Acknowledgement:}\\
We achnowledge financial support by the Deutsche Forschungsgemeinschaft
(DFG) in the framework of the priority programme SPP 1594 (Grant No.~HO
2231/8-2). PC thanks FOR 1394 for financial support during visit to
University of D\"usseldorf.

\newpage
%%%%%%%%%
%%%%%%%%%

\newpage

%%%%%%%%%
%%%%%%%%%
\newpage

%%%%%%%%%
%%%%%%%%%
\newpage
%%%%%%%%%
%%%%%%%%%
\newpage
%%%%%%%%%
%%%%%%%%%
\newpage
%%%%%%%%%
%%%%%%%%%
\newpage

%%%%%%%%%
%%%%%
\newpage
%%%%%
%%%%%%%%%
\newpage
%%%%%%%%%

\end{document}